\documentclass{appolb}
\usepackage{graphicx}
\usepackage{wasysym}
\usepackage{amssymb}
\usepackage{booktabs}
\usepackage{hyperref}

\begin{document}
\title{On the Charged Fragments Tagging in the ATLAS Detector during the 2025 Oxygen Campaign%
\thanks{Presented at XXXII Cracow Epiphany Conference on the recent results from Heavy Ion Physics, 13--17 January 2025, Krak\'ow, Poland.}%
}
\author{Weronika Sobie\'n
\address{Warsaw University, Faculty of Physics\\Ludwika Pasteura 5, 02-093 Warsaw, Poland.}
\\[3mm]
{Maciej Trzebi\'nski
\address{Institute of Nuclear Physics Polish Academy of Sciences\\ul. Radzikowskiego 152, 31-342 Krak\'ow, Poland.}
}
\\[3mm]
on behalf of the ATLAS Collaboration
}
\maketitle

\begin{abstract}
During the Summer of 2025 the LHC collided protons with oxygen, oxygen with oxygen and neon with neon. The ATLAS experiment recorded these data with its Forward Proton detectors (AFP) inserted on both: the proton and ion sides. This allows access to many interesting studies with scattered fragments being tagged. A few analysis ideas are presented followed by preliminary studies of what could be visible in the AFP.
\end{abstract}

\section{Introduction}

The proton--oxygen ($p$O) and oxygen--oxygen (OO) runs performed at the Large Hadron Collider (LHC) in June 2025 addressed a long-standing request from the astroparticle physics community \cite{oxygen_gen}. Data from such collisions is expected to provide essential input for modelling both the production of cosmic rays in astrophysical accelerators and their subsequent interactions with nuclei in the Earth's atmosphere \cite{QGSJET, angantyr, EPOS, Sibyll}. In addition, a dedicated neon--neon (NeNe) fill was carried out, enabling complementary studies of nuclear interactions in a different light-ion system \cite{ion_shape}. Collectively, these runs extend the LHC heavy-ion programme towards small and intermediate-sized systems, providing data that can put new constraints on QCD dynamics, nuclear structure and particle production mechanisms.

An important feature of nuclear collisions is the presence of \textit{spectator nucleons}, \textit{i.e.} nucleons that do not participate directly in the primary interaction. These spectators may form nuclear fragments that retain a large fraction of the beam momentum and are produced at very small angles with respect to the beam direction. The detection of such fragments provides valuable information on the collision geometry and on the breakup of the colliding nuclei \cite{ion_phys_ATLAS}.

A largely unexplored domain concerns measurements of very forward charged fragments \cite{fwd_models_1, fwd_models_2}. In nuclear collisions, such measurements provide direct sensitivity to diffractive and photon-induced processes, nuclear breakup channels and peripheral electromagnetic interactions. In particular, in OO collisions, forward fragment detection enables studies of nuclear excitation and fragmentation mechanisms.

Finally, if the identification of ion fragments is feasible, a wide range of measurements becomes accessible. In particular, cross-section measurements for specific fragment species would provide valuable constraints for Monte Carlo models and enable studies of nuclear fragmentation processes, such as beam transmutation \cite{transmutation} or $\alpha$-particle production.

The ATLAS experiment \cite{ATLAS} participated in this programme with extended forward measurement capabilities, \textit{i.e.} at high rapidities. The Zero Degree Calorimeter (ZDC) \cite{ZDC} was used to detect neutral fragments on the ion(s) side(s), while the ATLAS Forward Proton (AFP) \cite{AFP} detectors were inserted on both sides throughout the data taking. During the $p$O run, the LHCf detectors \cite{LHCf} were installed on the proton side, further extending the forward coverage. The LHCf performance was additionally enhanced by the use of all three ZDC hadronic modules.


\section{ATLAS Forward Proton Detectors}

The AFP system consists of four Roman pot stations, arranged symmetrically on both sides of the ATLAS interaction point. The stations located at 204~m are referred to as \textit{NEAR}, while those at 217~m are denoted as \textit{FAR}. Each station is equipped with 3D edgeless silicon tracking detectors (SiT), while the FAR stations additionally include a Time-of-Flight (ToF) system, which was switched off during the data-taking considered here.

The primary purpose of the SiT system is to provide precise reconstruction of particle trajectories, enabling the determination of their kinematic properties at the interaction point. Each station contains four detector planes, with each plane comprising $336 \times 80$ pixels of $50 \times 250~\mu$m$^2$ size and 230~$\mu$m thickness. The detectors are based on the FE-I4b readout chip \cite{FEI4} and are designed to be radiation-hard. The use of edgeless technology reduces the inactive region on the beam-facing side to approximately 100~$\mu$m. The expected spatial resolution is approximately 6~$\mu$m in the horizontal ($x$) direction and 30~$\mu$m in the vertical ($y$) direction \cite{AFP_TB}.

A proper configuration (\textit{tuning}) of the SiT detectors is essential for optimal data quality. In standard proton--proton operation, a typical Time-over-Threshold (ToT) setting corresponds to 10 at 20~ke$^-$. However, during the oxygen campaign, significantly larger charge deposits were expected due to the presence of heavier ion fragments. Therefore, a dedicated tuning configuration was adopted, with a threshold of 50~ke$^-$ at 6 ToT with intention to reach the maximal dynamic range of SiT sensors. This adjustment should broaden the AFP ion identification capability for boron, carbon and possibly nitrogen.

\section{Scattered Proton Measurements}

\begin{figure}[!htbp]
  \centering
  \begin{minipage}{0.49\textwidth}
    \centering
    \includegraphics[width=0.6\textwidth]{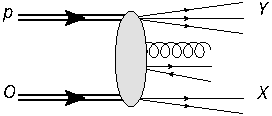}
  \end{minipage}
  \hfill
  \begin{minipage}{0.49\textwidth}
    \centering
    \includegraphics[width=0.6\textwidth]{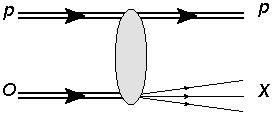}
  \end{minipage}
  \caption{Feynman diagram of non-diffractive (\textbf{left}) and diffractive (\textbf{right}) proton-oxygen interaction.}
  \label{diag_shower}
\end{figure}

One of the primary goals of the campaign was to provide input for the modelling of ultra-high-energy cosmic-ray air showers \cite{shower_sys_1, shower_sys_2}. Such data were provided by the LHC when a proton beam collided with an oxygen beam, resembling interactions of cosmic rays with atmospheric nuclei. These types of interactions can proceed via non-diffractive or diffractive mechanisms, as illustrated in Fig.~\ref{diag_shower}.

Using only the central ATLAS detector, the discrimination between these interaction classes is limited, as their signatures can appear similar in the central region. The inclusion of forward proton detectors gives a powerful handle by providing direct information about scattered proton.

\begin{figure}[!htbp]
  \centering
  \begin{minipage}{0.65\textwidth}
    \centering
    \includegraphics[width=1.0\textwidth]{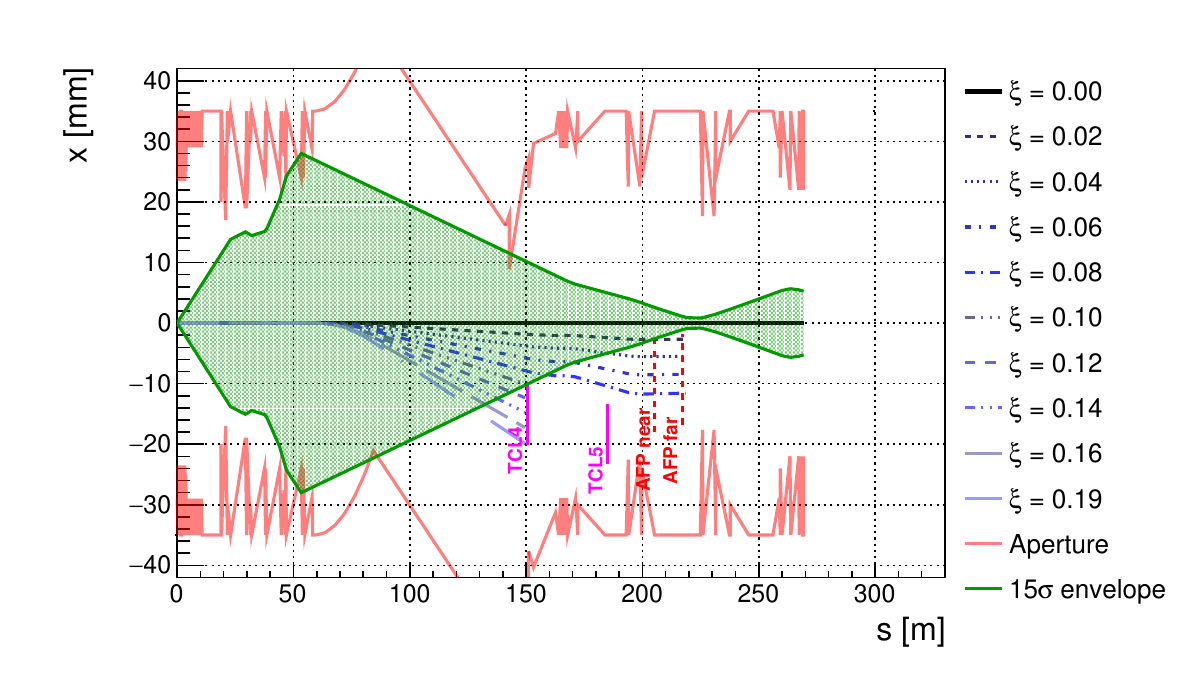}
  \end{minipage}
  \hfill  
  \begin{minipage}{0.33\textwidth}
    \centering
    \includegraphics[width=1.0\textwidth]{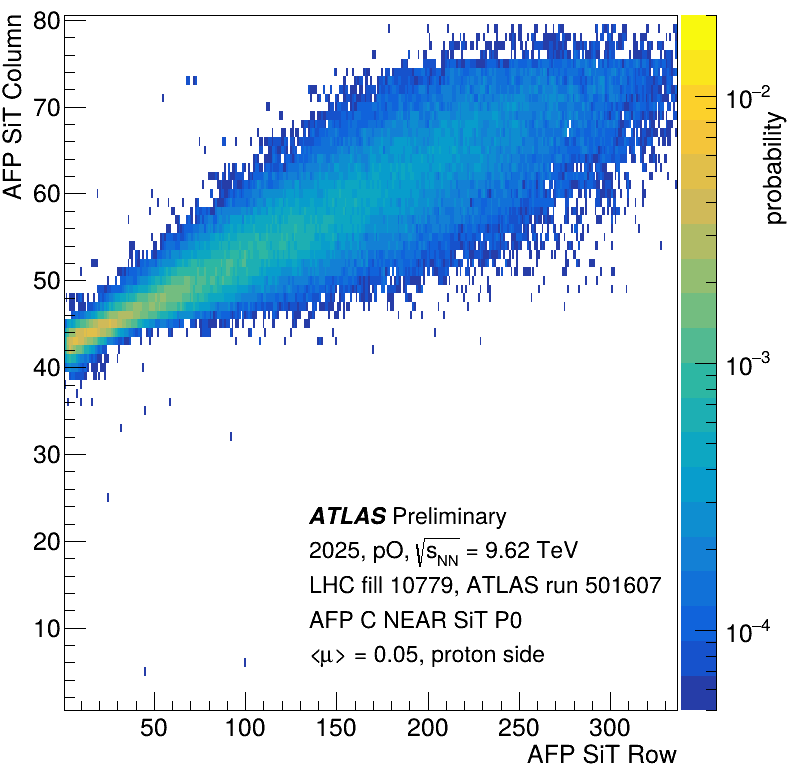}
  \end{minipage}
  \caption{\textbf{Left}: proton trajectories between ATLAS collision point and forward proton detectors. Various dotted and dashed lines represent situation in which proton lost a certain portion of its initial energy, $\xi = 1 - E_{proton}/E_{beam}$. Particles significantly diverging from the beam are stopped by the collimators: TCL4 and TCL5. \textbf{Right}: diffractive pattern registered in AFP station C NEAR SiT plane 0 on proton side during one of $pO$ runs. From \cite{FWD_public}.}
  \label{proton_pattern}
\end{figure}

The proton transport to the AFP detectors is governed by the LHC magnetic lattice, consisting of dipole and quadrupole magnets. These settings, referred to as the \textit{optics}, determines the mapping between the proton kinematics at the interaction point and its position in the detectors. Figure~\ref{proton_pattern} (left) illustrates proton trajectories for different values of the fractional energy loss, $\xi = 1 - E_{\mathrm{proton}}/E_{\mathrm{beam}}$.

The LHC optics convoluted with kinematics of the event leads to a characteristic \textit{diffractive pattern} in the AFP detectors. An example of such a pattern is shown in Fig.~\ref{proton_pattern} (right). The distributions of hit positions in detector coordinates --- pixel rows and columns --- reflect the underlying kinematics of the scattered protons. Protons with larger energy loss are deflected further from the beam and are therefore detected at larger horizontal positions. From the position measurements, the proton kinematics at the interaction point can be reconstructed \cite{unfolding_1, unfolding_2}.


\section{Ion Fragments}

In addition to proton measurements, the AFP detectors were also operated on the ion side to observe charged nuclear fragments. A key question concerns which fragment species can be detected within the AFP acceptance. As in the proton case, the determining factor is the LHC optics.

\begin{figure}[!htbp]
  \centering
  \includegraphics[width=0.41\textwidth]{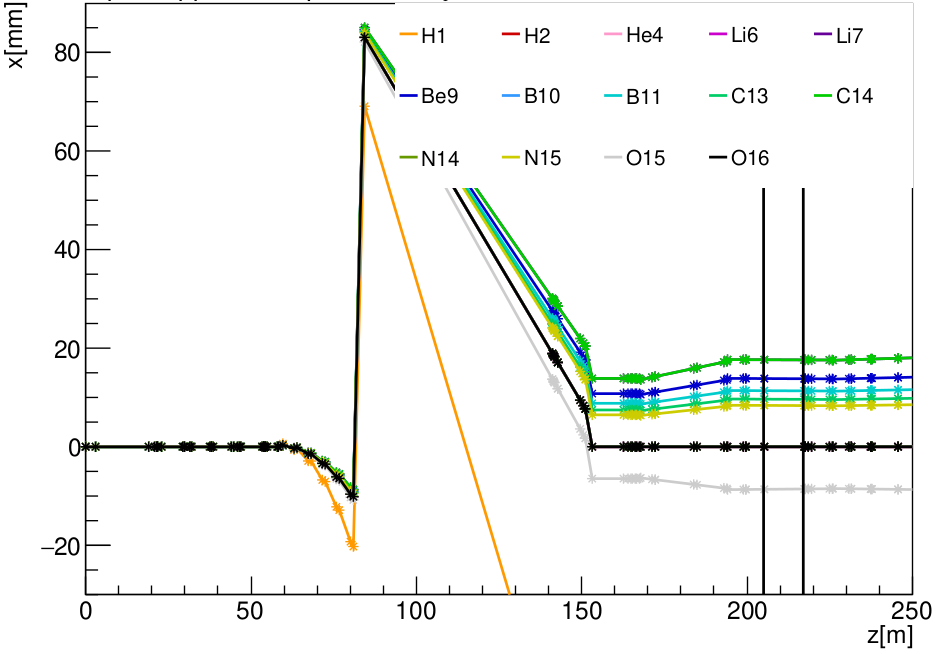}
  \includegraphics[width=0.55\textwidth]{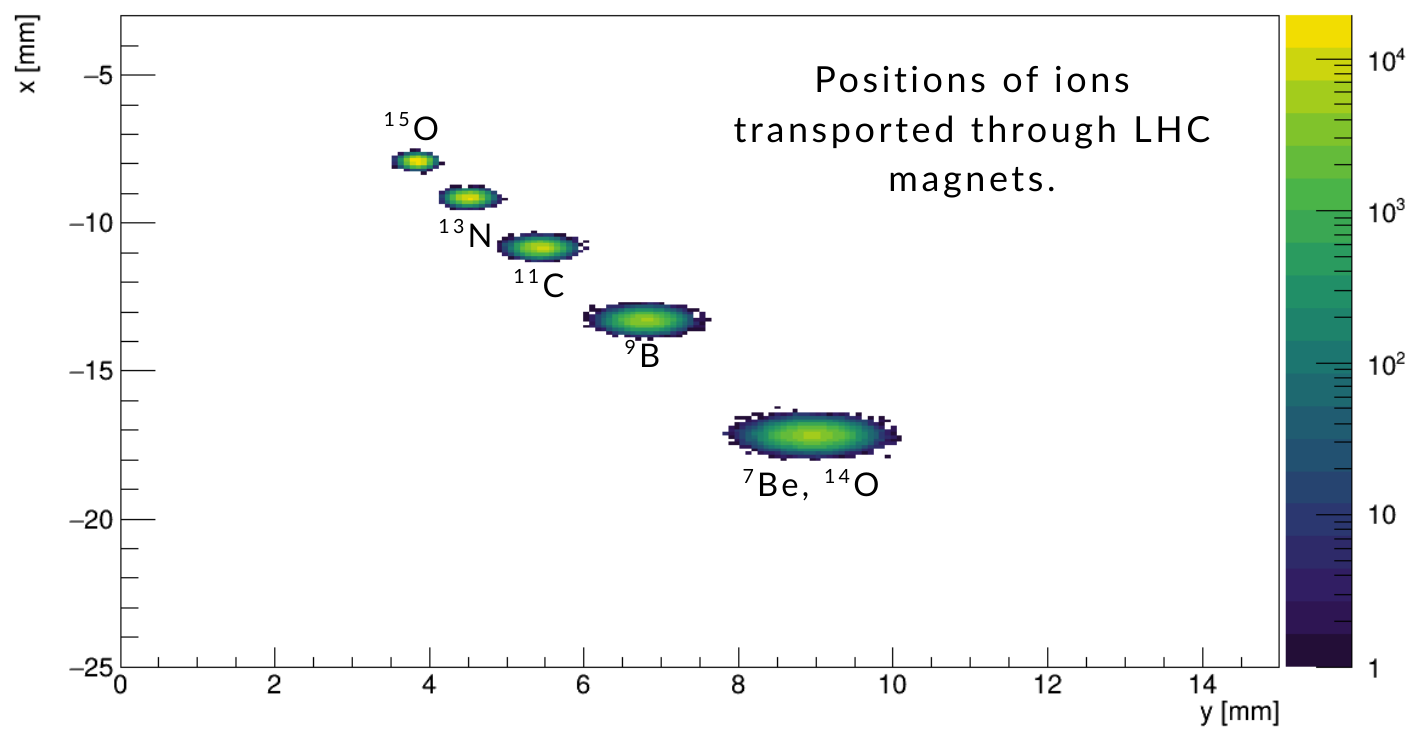}
  \caption{\textbf{Left}: trajectories of various ions between ATLAS collision point and forward proton detectors. The two black vertical lines indicate the locations of the AFP NEAR and FAR station, respectively. Trajectories for $^2$H, $^4$He, $^6$Li, $^{10}$B and $^{16}$O lie on top of each other. The 97 mm shift around $z = 80$ m reflects transition from a single beampipe present around ATLAS to two separated ones. \textbf{Right}: hit positions 205 m from ATLAS collision point. Transverse energy of ions was smeared accordingly to Fermi motion. Only fragments within the AFP geometric acceptance are drawn.}
  \label{ion_optics}
\end{figure}

The transport of ion fragments through the magnetic lattice depends on both their kinematics, atomic ($Z$) and mass ($A$) number. The LHC acts as a magnetic spectrometer -- see trajectories of selected ion species, Fig.~\ref{ion_optics} (left).

Simulations indicate that, among possible fragments, $^{7}$Be, $^{9}$B, $^{11}$C, $^{13}$N, $^{14}$O and $^{15}$O may fall within the AFP acceptance. The corresponding expected hit positions are shown in Fig.~\ref{ion_optics} (right). Particles with $Z/A = 0.5$, identical to that of $^{16}$O, follow trajectories close to the nominal beam orbit if they retain the full beam velocity. If fragments lose energy or acquire transverse momentum, they may be deflected enough to enter the detector acceptance.

\begin{figure}[!htbp]
  \centering
  \includegraphics[width=0.32\textwidth]{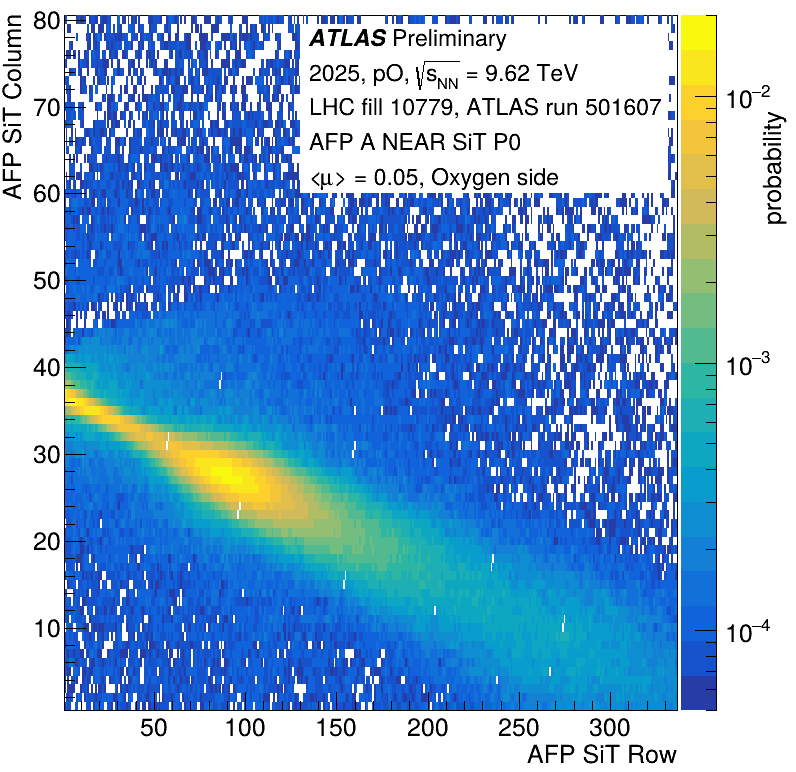}
  \includegraphics[width=0.32\textwidth]{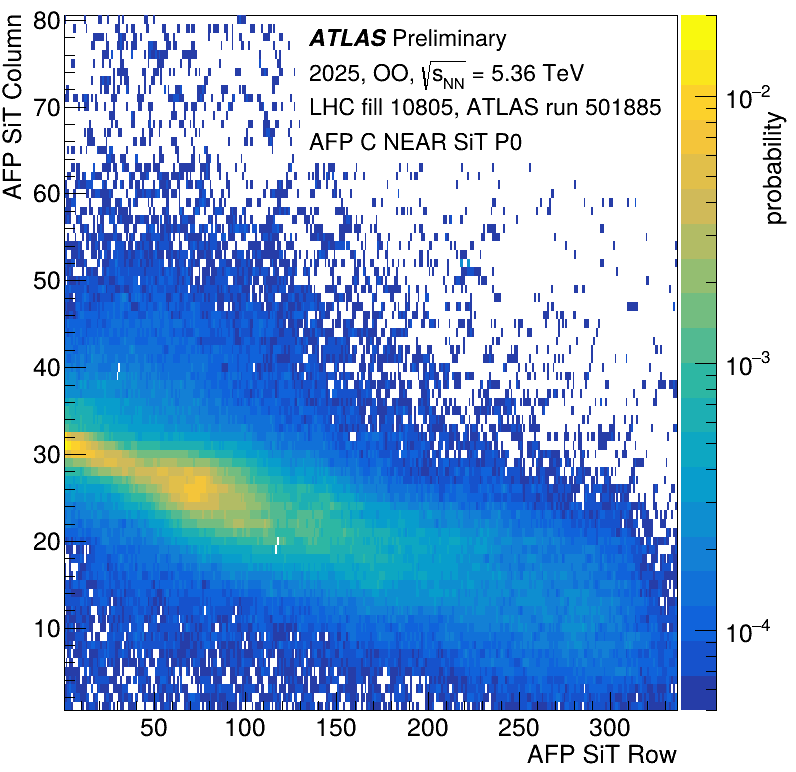}
  \includegraphics[width=0.32\textwidth]{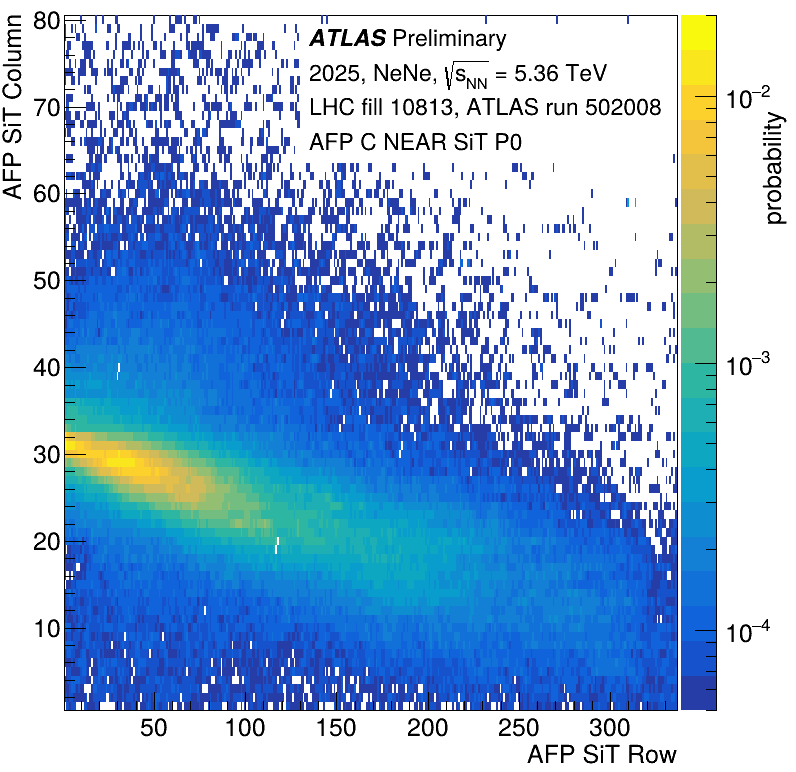}
  \caption{Hitmaps registered by the AFP detectors during proton-oxygen (\textbf{left}), oxygen-oxygen (\textbf{middle}) and neon-neon (\textbf{right}) collisions. From \cite{FWD_public}.}
  \label{hitmaps}
\end{figure}

Hitmaps recorded during $p$O, OO and NeNe data taking are presented in Fig.~\ref{hitmaps}. In contrast to the proton side, the ion side exhibits a more complex structure. In addition to a continuous distribution, potentially associated with fragments having $Z/A \approx 0.5$, localized enhancements are observed at specific detector positions. These features may indicate contributions from distinct fragment species.

\section{Summary}
The 2025 light-ion campaign at the LHC extends the heavy-ion programme to small and intermediate-sized systems and provides important input for both QCD studies and cosmic-ray physics. In particular, measurements in the very forward region may constrain particle production mechanisms relevant to extensive air showers.

The ATLAS experiment contributed to this programme using its forward instrumentation, combining the ZDC for neutral fragments and the AFP detectors for charged particles. On the proton side, AFP enables the identification of diffractive and photon-induced processes through the detection of intact scattered protons. On the ion side, it provides sensitivity to charged nuclear fragments.

Preliminary observations indicate interesting structures in the ion-side hit distributions, suggesting presence of various fragments. If confirmed, the capability to differentiate ion fragments would open new avenues for studying nuclear breakup and for constraining models of hadronic interactions.

\section*{Acknowledgements}
The work of MT and WS was partially supported by the Polish National Science Centre (project no. UMO-2019/34/E/ST2/00393). Copyright 2026 CERN for the benefit of the ATLAS Collaboration. CC-BY-4.0 license.



\begin{thebibliography}{99}
\bibitem{oxygen_gen}
J.~Brewer \textit{et al.}, CERN-TH-2021-028, \href{https://arxiv.org/abs/2103.01939}{arXiv:2103.01939}.

\bibitem{QGSJET}
S.~Ostapchenko, \href{https://arxiv.org/abs/2208.05889}{arXiv:2208.05889}.

\bibitem{angantyr}
Ch.~Bierlich \textit{et al.}, \href{https://arxiv.org/abs/1806.10820}{JHEP \textbf{10} (2018) 134}.

\bibitem{EPOS}
T.~Pierog \textit{et al.}, \href{https://arxiv.org/abs/1306.0121}{arXiv:1306.0121}.

\bibitem{Sibyll}
F.~Riehn \textit{et al.}, \href{https://doi.org/10.1103/PhysRevD.102.063002}{Phys.\ Rev.\ D \textbf{102} (2020) 063002}.

\bibitem{ion_shape}
ATLAS Collaboration, \href{https://arxiv.org/abs/2509.05171}{arXiv:2509.05171}.

\bibitem{ion_phys_ATLAS}
I.~Grabowska-Bold, \href{https://doi.org/10.5506/APhysPolBSupp.18.5-A10}{Acta Phys.\ Pol.\ B Proc.\ Suppl.\ \textbf{18} (2025) 5-A10}.

\bibitem{fwd_models_1}
S.~Ostapchenko, \href{https://arxiv.org/abs/1612.09461}{arXiv:1612.09461}.

\bibitem{fwd_models_2}
M.~Tasevsky, \href{https://arxiv.org/abs/1802.02818}{arXiv:1802.02818}.

\bibitem{transmutation}
G.~Nijs and W.~van~der~Schee, \href{https://arxiv.org/abs/2507.01659}{arXiv:2507.01659}.

\bibitem{ATLAS}
ATLAS Collaboration, \href{https://doi.org/10.1088/1748-0221/3/08/S08003}{JINST \textbf{3} (2008) S08003}.

\bibitem{ZDC}
G.~Avoni \textit{et al.}, \href{https://arxiv.org/abs/2509.05948}{JINST \textbf{20} (2025) P11021}.

\bibitem{AFP}
ATLAS Collaboration, \href{https://cds.cern.ch/record/2017378?}{CERN-LHCC-2015-009, ATLAS-TDR-024}.

\bibitem{LHCf}
LHCf Collaboration, \href{https://doi.org/10.1088/1748-0221/3/08/S08006}{JINST \textbf{3} (2008) S08006}.


\bibitem{FEI4}
M.~Garcia-Sciveres \textit{et al.}, \href{https://doi.org/10.1016/j.nima.2010.04.101}{Nucl.\ Instrum.\ Meth.\ A \textbf{636} (2011) S155}.

\bibitem{AFP_TB}
J.~Lange \textit{et al.}, \href{https://doi.org/10.48550/arXiv.1608.01485}{JINST \textbf{11} (2016) P09005}.

\bibitem{shower_sys_1}
R.~Parsons \textit{et al.}, \href{https://doi.org/10.48550/arXiv.1102.4603}{Astropart.\ Phys.\ \textbf{34} (2011) 832}.

\bibitem{shower_sys_2}
R.~D.~Parsons and H.~Schoorlemmer, \href{https://doi.org/10.1103/PhysRevD.100.023010}{Phys.\ Rev.\ D \textbf{100} (2019) 023010}.

\bibitem{FWD_public}
ATLAS Collaboration, \href{https://twiki.cern.ch/twiki/bin/view/AtlasPublic/ForwardDetPublicResults}{Forward Detector Public Results}.

\bibitem{unfolding_1}
R.~Staszewski and J.~Chwastowski, \href{https://doi.org/10.1016/j.nima.2009.08.023}{Nucl.\ Instrum.\ Meth.\ A \textbf{609} (2009) 136}.

\bibitem{unfolding_2}
M.~Trzebinski \textit{et al.}, \href{https://doi.org/10.5402/2012/491460}{ISRN High Energy Phys.\ \textbf{2012} (2012) 491460}.




\end{thebibliography}
\end{document}